# Effect of AlN Seed Layer on Crystallographic Characterization of Piezoelectric AlN

Kaitlin M. Howell [a) 1], Waqas Bashir[1], Annalisa de Pastina[1], Ramin Matloub[2], Paul[2] Muralt, Luis G. Villanueva[b) 1]

[1]Advanced NEMS Group, EPFL, Station 9, Bâtiment MED, Lausanne Switzerland

[2]Electroceramic Thin Films Group, EPFL, SCI-STI-PM, Station 12, Bâtiment MXD, 1015 Lausanne Switzerland

[a)] Electronic mail: kaitlin.howell@epfl.ch

[b)] Electronic mail: guillermo.villanueva@epfl.ch

Ultrathin aluminum nitride (AlN) films are of great interest for integration into nanoelectromechanical systems for actuation and sensing. Given the direct relationship between crystallographic texture and piezoelectric response, x-ray diffraction has become an important metrology step. However, signals from layers deposited below the piezoelectric (PZE) AlN thin film may skew the crystallographic analysis and give misleading results. In this work, we compare the use of a Ti or AlN seed layer on the crystallographic quality of PZE AlN. We also analyze the influence of several AlN seed layer thicknesses on the rocking curve FWHM of PZE AlN and demonstrate an larger effect of the AlN seed layer on the θ-2θ AlN <0002> crystallographic peak for increasing AlN seed layer thickness.

# I. INTRODUCTION

Micro/nanoelectromechanical systems (M/NEMS) have been an area of intense research in the past decades due to their high sensitivity, low fabrication cost, and fast response as well as availability of measurement modes and wide spectrum of measurable phenomena [1,2]. It is well known that miniaturization has several positive effects, such as improving the limit of detection in sensing mass [3] or gas [4] and enabling mechanics-based logic to become a viable alternative to metal-oxide-semi-conductor (MOS) transistors [5,6]. However, miniaturization comes at the cost of greater difficulty in fabrication and integration. To effectively actuate and detect the motion of M/NEMS, piezoelectricity has become highly utilized due to its low power consumption, scalability[7], extreme linearity and easy integration [8] in comparison with optical [9], electrostatic [10] or magnetomotive [11] transduction techniques.

Several piezoelectric materials have been explored for integration in M/NEMS, including lead zirconium titanate (PZT) [12], zinc oxide (ZnO) [13] and aluminum nitride (AlN) [14]. AlN is one of the most commonly used materials due to its low loss tangent (both dielectric and mechanical), structural and chemical stability. AlN thin films can be deposited via several methods, including molecular beam epitaxy (MO-CVD) [15], pulsed laser deposition [16] and reactive sputtering [17]. The latter is a standard microfabrication technique in both industry and research and enables the growth of highly-textured thin films with low thermal budgets and better residual stress control, thus compatible with CMOS processing [17]. As a consequence, AlN-sputtered thin films are commercially used in thin film bulk-wave acoustic resonator (TFBAR) filters[18] and have been reported at the academic research stage in contour mode resonators (CMRs) [19], switches [20,21], suspended microchannel resonators (SMRs) [22] and accelerometers [23].

To optimize the piezoelectric transduction efficiency in flexural resonators, it is widely known that one should use as thin as possible piezoelectric layer still keeping a large-enough breakdown voltage[2,8,24,25]. However, AlN ultra-thin films with thicknesses less than 200 nm have been shown to have worse crystalline quality than their thicker counterparts [26]. To improve the crystalline quality of AlN ultra-thin films, some investigations have been made, such as using multiple deposition steps during reactive sputtering [24] and adding annealing during AlN sputtering deposition steps [27].

In order to use AlN to actuate and detect the motion of M/NEMS, AlN must be sandwiched by metal electrodes. The most typically used are aluminum (Al), due to its low density, molybdenum (Mo), due to its high acoustic velocity, and platinum (Pt), since it has been shown that Pt gives the best growth conditions for highly textured AlN thin films [28]. In order to improve the surface quality and adhesion of said bottom metal, a thin seed layer has been repeatedly used between the electrode and the substrate. Two typical examples stand out: Ti and AlN. It has been shown that a thin (~10 nm) Ti layer serves as a good adhesion promoter for the metal electrode, e.g. Pt, to the underlying substrate [26]. AlN has been utilized in a number of works with piezoelectriclayer thicknesses of or below 100nm[8,20,24,29]. However, little has been said on the effect of seed layers on the crystallization of the thin films grown above it.

In this work, we investigate the effect of a seed layer on the crystalline and piezoelectric properties of the piezoelectric layer of AlN (from now on the "PZE layer") that is sandwiched between Pt electrodes. We first compare Ti versus AlN as a seed layer in the growth of highly textured AlN thin films through x-ray diffraction (XRD) analysis. After, PZE layers grown on top of various AlN seed layer thicknesses are prepared.

Through characterization of the crystallinity and piezoelectric response, an apparent loss of crystallographic quality is revealed despite a slightly improved piezoelectric response compared to state of the art. This apparent contradiction (with reported values) is explained by demonstrating that the XRD measurements are a convolution of two peaks, one from the AlN seed layer and one from the PZE layer, which leads to a broadening of the rocking curve full width half maximum (FWHM).

## II. EXPERIMENTAL DETAILS

### A. Thin Film Deposition

All thin film depositions are attained with a DC magnetron sputtering tool (Pfeiffer Spider 600) on 380 μm thick <100> silicon wafers with 1 um wet $SiO_2$. In order to clean any possible residues residing on the substrates, a radio frequency (RF) etch step of 15 seconds is performed within the tool directly before deposition of the thin films. For the comparison of Ti versus AlN as a seed layer, a full stack of four thin films is deposited, consisting of the seed layer (15 nm of Ti or AlN), 25 nm of Pt as a bottom electrode, 50 nm of AlN for the PZE layer and 25 nm of Pt as the top electrode. To measure the influence of the AlN seed layer thickness, substrates are prepared with only AlN seed layers of various thicknesses (15-100 nm) and compared to substrates with full stacks and the same seed layer thicknesses. All AlN depositions are done with a 99.9995% Al target at a chamber temperature of 350°C. Base pressure of the sputtering chamber during deposition is $2\times10^{-3}$ mbar. Gas flows introduced during sputtering are 10 sccm of Ar and 40 sccm of $N_2$.

### B. Crystallographic and Piezoelectric Characterization

A XRD system (D8 Discover, Bruker, Germany) with a resolution of 0.001° is used to take θ-2θ scans over a large range, including the Si <400> peak and the <0002> AlN peak, the later also being characterized by rocking curves. The piezoelectric quality of the films are characterized with a double laser interferometric setup, described elsewhere [30] and compared to finite element analysis to correlate electrode size with the apparent piezoelectric response[31].

## C. Test Devices Fabrication

The fabrication process for the test devices used in the piezoelectric response measurements is shown in Figure 1. After the deposition of the full stack (Figure 1 (A)), a photolithography step is made to pattern the top two layers (Figure 1 (B)). The top Pt layer is etched in an ICP tool with $Cl_2$/Ar chemistry. The PZE layer is then selectively wet etched in phosphoric acid (Figure 1 (C)). A second photolithography step is used to define the bottom two layers (Figure 1 (D)); $Cl_2$/Ar chemistry is used to etch both the bottom Pt layer and AlN seed layer (Figure 1 (E)).

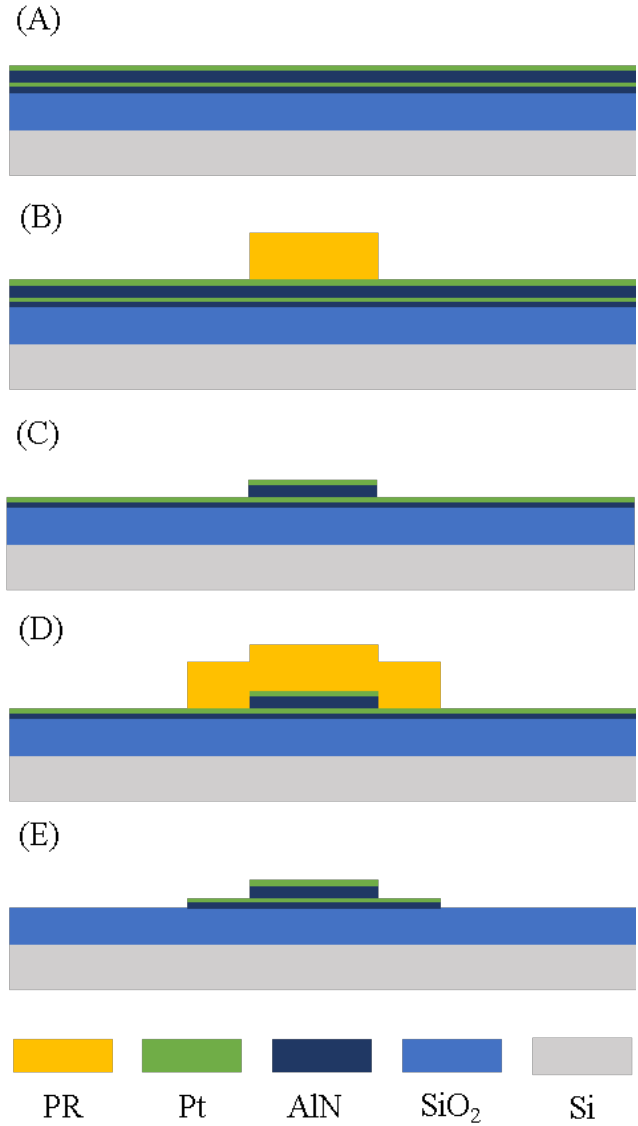

FIG. 1. (Color Online) Process flow for fabrication of electrodes for x-ray diffraction and spiezoelectric characterization. A) Deposition of full stack by magnetron sputtering. B) Photolithography to define top electrodes and PZE layer. C) Dry and wet etching of the top electrodes and PZE layer, respectively. D) Photolithography to define bottom electrodes and AlN seed layer. E) Dry etching of bottom electrodes and AlN seed layer. Cross sections are not to scale.

## III. EXPERIMENTAL RESULTS

We first measure θ-2θ and rocking curves on full stacks with either a Ti or AlN seed layer. The FWHM of the rocking curves are calculated through a Gaussian fitting: a Ti seed layer provides a full stack rocking curve FWHM of approximately 2.6°, while a FWHM of 2.0° is found when using an AlN seed layer of the same thickness, i.e. 15 nm. Based on work by F. Martin et al.[26], who found a correlation between the FWHM of rocking curves of the AlN <0002> crystallographic peak and measured piezoelectric coefficients, we deduce that AlN as a seed layer material yields growth of a higher quality PZE layer compared to Ti. As a consequence, we decide to focus our efforts on using AlN as a seed layer.

To study the influence of the AlN seed layer thickness on the piezoelectric quality of the PZE layer, four different AlN seed layer thicknesses (15, 25, 50 and 100 nm) are deposited on different wafers and compared to deposited full stacks with the same respective seed layer thicknesses. The FWHM of the rocking curves measured for both the seed layers and full stacks are shown in Figure 2. As the seed layer thickness increases, the FWHM of its measured rocking curve significantly decreases, signaling an improvement in crystalline quality. However, a small increase is seen in the FWHM of the full stack as the seed layer thickness increases, a seemingly surprising result since it is expected that improved crystallinity of underlying layers should improve the crystallinity of the PZE layer. In addition, based on the work from Martin et al. [26], one could go further and deduce that the piezoelectric coefficient for PZE layers with a thicker seed layer are worse than those with a thinner seed layer.

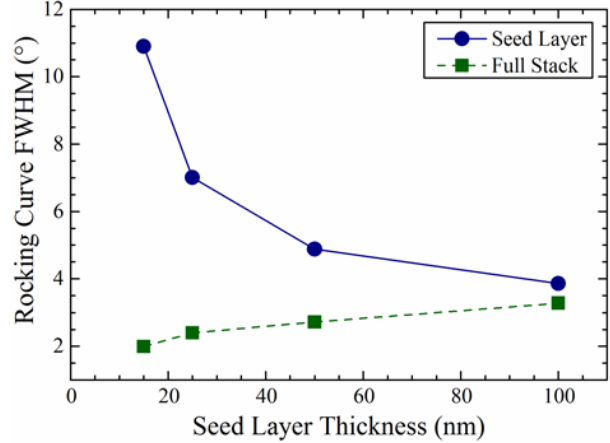

FIG. 2. (Color Online) Measured rocking curve FWHM of AlN seed layer only (Seed Layer) and full stacks with four different seed layer thicknesses. Note the decrease in FWHM for the AlN seed layer due to an increase in crystallinity as the layer thickness increases. However, there is an increase in FWHM for the full stack when the AlN seed layer thickness increases. Lines in graph are for visual purposes, not trends.

In order to check this latter point, we pattern test devices with several different radii on samples with the full stack of deposited layers to measure the piezoelectric response. In Figure 3, we can see that the piezoelectric response, shown as vertical displacement per unit volt, depends on the electrode radius. This has been reported in the past [32] and that can be reproduced by Finite Element Modelling (FEM) [33] (see Figure 3 inset). We find that all devices, no matter the seed-layer thickness, fabricated in this work show an average $d_{33,f}$ of $3.51 \pm 0.17$ pm/V, which is slightly better than thicker films of AlN ($d_{33,f} = 3.4 \pm 0.17$ pm/V)[17]. Therefore, based on results shown in Figure 2 and 3, the piezoelectric response is independent of the measured crystallinity for the full stack. This might seem, at first sight, to contradict the work by Martin et al. [26]; however, we will demonstrate an explanation for this contradiction by further analyzing the x-ray diffraction measurements.

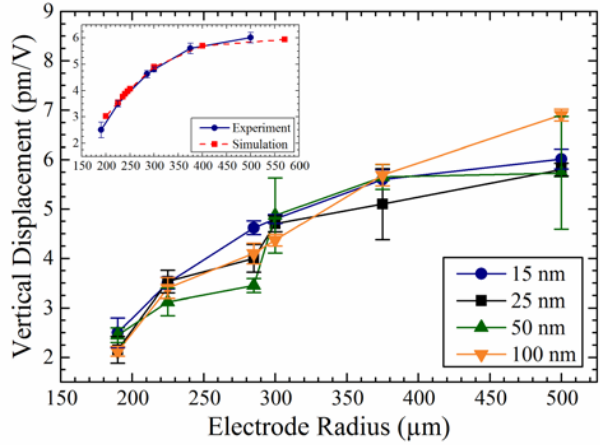

FIG. 3. (Color Online) Comparison of measured piezoelectric response versus electrode radius for full stacks with 4 different AlN seed layer thicknesses. The measured responses with each AlN seed layer thickness do not significantly differ from each other as electrode radius increases. Inset demonstrates the overlap between FEM of the response from a full stack with a 15 nm AlN seed layer and complementary experimental data for different electrode radii. Lines in graph are for visual purposes, not trends.

## IV. Gaussian Analysis

The discrepancy between XRD measurements and piezoelectric response can be explained by taking into account that the XRD measurement of the overall stack returns signals not only from the PZE layer, but also the AlN seed layer, due to the fact that the X-rays penetrate and interact with both layers. Therefore, a Gaussian analysis in MATLAB® is completed on the θ-2θ data of both the seed layer only and full stack samples to determine the contribution of both layers to the full stack <0002> crystallographic peak.

First, a 1-term Gaussian with a linear background subtraction is fitted to the AlN seed layer θ-2θ <0002> crystallographic peak to find the peak center ($A_2$) and its standard deviation ($A_3$). These values are then applied to a 2-term Gaussian with a linear background that is fitted to the full stack <0002> peak, which is shown in Eq (1):

$$A_1 \exp(-(x-A_2)^2/2A_3^2) + B_1 \exp(-(x-B_2)^2/2B_3^2) + C_1 x + C_2 \qquad (1)$$

Where the first exponential term corresponds to the AlN seed layer, the second to the PZE layer and $C_1$ and $C_2$ correspond to a linear background. As discussed above, variables $A_2$ and $A_3$ were calculated through Gaussian fittings of the AlN seed layer-only samples and applied in Eq. (1). All other variables were left free to be fitted. The results of the 2-term Gaussian fitting are shown in Table 1. The Gaussian fits for each full stack demonstrated a high adjusted $R^2$. In Table 1, the ratio of the integrated areas under each Gaussian term is found to be directly related to the thickness volume fraction ($t_{seed}/(t_{seed} + t_{PZE})$).

TABLE I. Results of Gaussian analysis of full stacks with four different AlN seed layer thicknesses ($t_{seed,AlN}$). The AlN seed layer/PZE layer peak area ratio demonstrates the relative significance of the AlN seed layer in the total measured <0002> crystallographic peak. All fits demonstrated a high adjusted $R^2$.

| $t_{seed}$ (nm) | Seed Layer Volume Fraction (%) | Seed Layer/PZE Layer Peak Area Ratio | Peak Area/Volume Fraction | Gaussian Fit AdjR$^2$ |
|---|---|---|---|---|
| 15 | 23 | 0.67 | 2.91 | 0.99 |
| 25 | 33 | 1.07 | 3.22 | 0.99 |
| 50 | 50 | 1.59 | 3.19 | 0.99 |
| 100 | 67 | 1.83 | 2.74 | 0.99 |

The results of the 2-term Gaussian fitting for each full stack are plotted against the raw data in Figure 4 (A-D). Each Gaussian term (and therefore the signal from each AlN layer) to the 2-term Gaussian fitting are also graphed to demonstrate their respective contributions. The conclusion that can be drawn from these results is that if AlN is used as a seed layer in the manner described above, one must ensure that the seed layer is much thinner than the PZE layer to minimize discrepancies between XRD measurements and piezoelectric response.

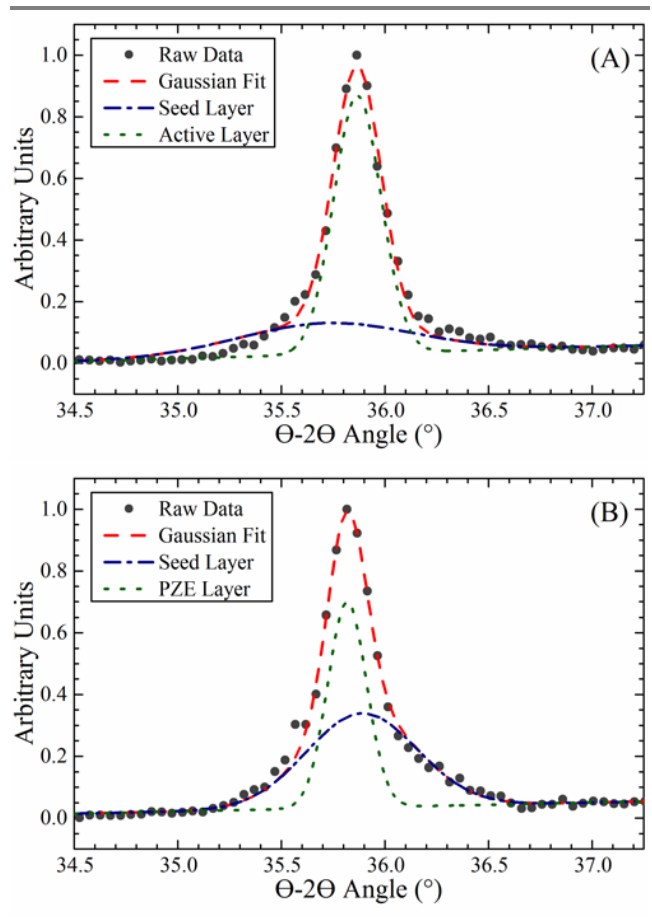

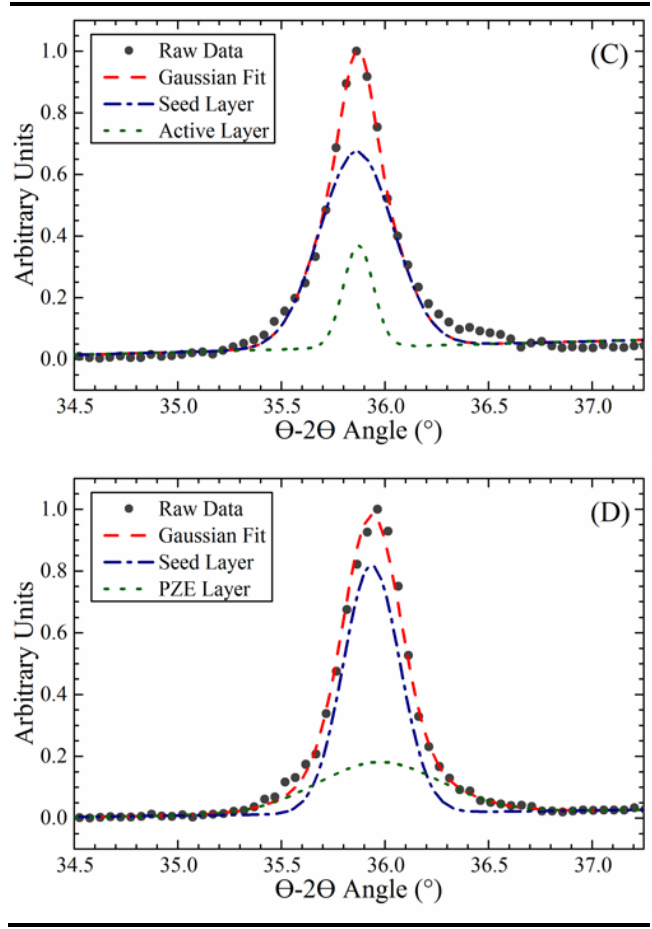

FIG. 4. (Color Online) Results of 2-term Gaussian fit of full stack θ-2θ <0002> crystallographic peak for 15, 25, 50 and 100 nm AlN seed layer thicknesses (A-D graphs respectively). As the thickness of the AlN seed layer increases, its influence on the full stack's <0002> crystallographic peak increases.

## V. SUMMARY AND CONCLUSIONS

In this paper, we report on studies of AlN as an alternative seed layer to Ti for enhanced growth of AlN thin films. First, the influence of Ti and AlN on the crystallinity of AlN PZE layers is measured, demonstrating that AlN is a suitable or better alternative to Ti. Then, AlN PZE layers were grown in a full stack with four AlN seed layer thicknesses. While piezoelectric response measurements are comparable to state of the art, rocking curve measurements demonstrate an influence of the AlN seed layer on the full stack results. This discrepancy between rocking curve FWHM and piezoelectric response lead to a Gaussian analysis of θ-2θ curves of AlN seed and full stack samples with the same seed layer thickness. It is demonstrated that if the AlN seed layer thickness is close or larger than the PZE layer thickness, its contribution to the measured <0002> crystallographic peak of AlN can increase, enlarging the measured full stack peak. However, piezoelectric response is not influenced by AlN seed layer thickness. Therefore, if rocking curve analysis is used to validate AlN thin films grown on a AlN seed layer, either the ratio of AlN seed/PZE layer thickness must be low or the analysis must be assumed to be skewed towards worse results than in reality.

Acknowledgements: The authors acknowledge financial support from the Swiss National Science Foundation (SNSF) under grant PP00P2_170590. All samples were fabricated in the Center of MicroNanoTechnology (CMi) at EPFL.